\documentclass[sigconf]{acmart}
\AtBeginDocument{%
  \providecommand\BibTeX{{%
    \normalfont B\kern-0.5em{\scshape i\kern-0.25em b}\kern-0.8em\TeX}}}

\setcopyright{acmcopyright}
\copyrightyear{2018}
\acmYear{2018}
\acmDOI{10.1145/1122445.1122456}

\acmConference[Woodstock '18]{Woodstock '18: ACM Symposium on Neural
  Gaze Detection}{June 03--05, 2018}{Woodstock, NY}
\acmBooktitle{Woodstock '18: ACM Symposium on Neural Gaze Detection,
  June 03--05, 2018, Woodstock, NY}
\acmPrice{15.00}
\acmISBN{978-1-4503-XXXX-X/18/06}

\usepackage{amsmath}
\usepackage{amsthm}
\usepackage{booktabs}
\usepackage{multirow}

\makeatletter
\newif\if@restonecol
\makeatother

\usepackage[ruled,linesnumbered]{algorithm2e}

\newcommand{\hv}{\mathbf{h}}

\newcommand{\Ev}{\mathbf{E}}

\newcommand{\argmin}{\operatornamewithlimits{argmin}}
\newcommand{\argTopk}{\operatornamewithlimits{argTopk}}

\begin{document}

\title{WSLRec: Weakly Supervised Learning for Neural Sequential Recommendation Models}

\author{Jingwei Zhuo, Bin Liu, Xiang Li, Han Zhu, Xiaoqiang Zhu}
\email{{zjw169463, zhuoli.lb, yushi.lx, zhuhan.zh, xiaoqiang.zxq}@alibaba-inc.com}
\affiliation{%
  \institution{Alibaba Group}
  \city{Beijing}
  \country{China}
}

\renewcommand{\shortauthors}{Trovato and Tobin, et al.}
\newcommand{\bin}[1]{\textcolor[rgb]{1,0,0}{(Bin: #1)}}

\begin{abstract}

Learning the user-item relevance hidden in implicit feedback data plays an important role in modern recommender systems.
Neural sequential recommendation models, which formulates learning the user-item relevance as a sequential classification problem to distinguish items in future behaviors from others based on the user's historical behaviors, have attracted a lot of interest in both industry and academic due to their substantial practical value.
Though achieving many practical successes, we argue that the intrinsic {\bf incompleteness} and {\bf inaccuracy} of user behaviors in implicit feedback data is ignored and conduct preliminary experiments for supporting our claims.
Motivated by the observation that model-free methods like behavioral retargeting (BR) and item-based collaborative filtering (ItemCF) hit different parts of the user-item relevance compared to neural sequential recommendation models, we propose a novel model-agnostic training approach called WSLRec, which adopts a three-stage framework: pre-training, top-$k$ mining, and fine-tuning.
WSLRec resolves the incompleteness problem by pre-training models on extra weak supervisions from model-free methods like BR and ItemCF, while resolves the inaccuracy problem by leveraging the top-$k$ mining to screen out reliable user-item relevance from weak supervisions for fine-tuning.
Experiments on two benchmark datasets and online A/B tests verify the rationality of our claims and demonstrate the effectiveness of WSLRec.
%The proposed framework mainly has two modules. 
%The first one is weakly supervised preference signal introduction module that aims to mine potential but not observed user preference. 
%A following preference distinguish module is used to identify those truly useful signals from the mined weakly supervised data and user feedback data, to train the recommendation model. 
%We will show that the proposed training framework is effective when adapting with most of the existing recommendation models, and can significantly improve the recommendation performance. 

% Recommendation is usually formulated as a classification problem, where a model is trained to distinguish the items a user may be interested in from the rest ones, mainly based on the user's behavior history, profile and other auxiliary information.
% Retrieving a subset from a large candidate set forms the core of modern recommender systems.
% Implicit feedback data prevails in modern recommender systems.
% Most of them formulates the retrieval problems as a classification problem to distinguish positive implicit feedback and negative ones.
% We argue that this formulation has two drawbacks:
    
\end{abstract}




\maketitle

\section{Introduction}

Recommending users with relevant items which may match their interest and thus trigger their positive reaction forms the core of recommender systems.
In real-world scenarios, the user interest usually hides in users' historical behaviors and is intrinsically dynamic and evolving, which poses a challenge to the quality of recommendations.
Recent years have witnessed a trend of using neural sequential recommendation models~\cite{covington2016deep,cheng2016wide,kang2018self,zhu2018learning,li2019multi,cen2020controllable} for solving this problem.
Compared to collaborative filtering and matrix factorization~\cite{sarwar2001item,koren2009matrix}, neural sequential recommendation models take the sequential dynamics of user behaviors into account directly by leveraging deep neural networks (DNNs) for learning rich representations on behavior sequences and modeling the complex relevance relations between users and items.

While much effort has been paid for developing advanced model architecture, how to train these models has received much less attention so far.
The ``standard'' way usually adopts a sequential classification formulation and trains DNNs to distinguish relevant items (i.e., positive labels) from irrelevant ones (i.e., negative labels) given the users' historical behaviors, where only the next immediate or next several items in future behaviors are regarded as relevant items and the rest in the corpus is regarded as irrelevant ones.
We argue that this purely supervised learning approach ignores the intrinsic {\bf incompleteness} and {\bf inaccuracy} of user behaviors in implicit feedback data, which is almost the default setting for recommender systems since the data with explicit user-item relevance is hard to collect~\cite{joachims2005accurately,hu2008collaborative}.
In practice, since only a small subset of items are exposed to users and users' behaviors are limited in this subset, the observed behaviors only reflect parts of the relevance relations between users and items, which results in that the trained models get stuck in local optimal and prevents the recommender system from exploring novel user interest. However, if we try to involve more supervision signals such as user's long-term future behaviors to supplement and extend the relevance information, the accurateness of those additional signals is not guaranteed.

To solve the incompleteness problem, we resort to other signals besides pure implicit feedback from users.
We consider model-free methods like behavioral retargeting (BR)~\cite{yan2009much} and item-based collaborative filtering (ItemCF)~\cite{sarwar2001item}, which are widely used in e-commerce recommender systems. 
These methods are different from future behavior prediction in neural sequential recommendation models: BR recommends a user with items that are in the historical behaviors and ItemCF recommends items similar to the historical behaviors where the similarity between items is measured globally. 
Our initial experiment in Table~\ref{tab:demo2} demonstrates that these methods hit different parts of the user-item relevance compared to neural sequential recommendation models. 
However, these additional signals are quite noisy, and taking them directly as relevant items in model training will exacerbate the inaccuracy problem.

For resolving these problems, we propose a novel model-agnostic training approach called {\bf W}eakly {\bf S}upervised {\bf L}earning for Neural Sequential {\bf Rec}ommendation Models (WSLRec).
By regarding them accompanying with those in future behaviors uniformly as weak supervisions, WSLRec adopts a three-stage framework: pre-training, top-$k$ mining, and fine-tuning.
In the pre-training stage, weak supervision is used to train the model directly.
Then, the top-$k$ item set with $k$-largest prediction scores according to the pre-trained model is generated as a filter for mining relevant items from weak supervisions.
Finally, the model is fine-tuned on these mined items.
It's worth mentioning that sources of supervision signals other than BR and ItemCF are also feasible to be involved in WSLRec, while we only use them in this paper for verification.

To the best of our knowledge, the drawbacks of such a ``standard'' training approach are also discussed in previous work~\cite{ma2020disentangled,zhou2020s3,wang2020denoising}.
\citet{ma2020disentangled} proposes to mine extra latent supervision signals from longer-term future behaviors based on a specific model architecture for encoding disentangled representations on historical behaviors,
$\mathrm{S}^3$-Rec~\cite{zhou2020s3} focuses on the incompleteness problem and proposes to leverage the intrinsic correlation among data incorporated with item attributes to derive extra self-supervision signals according to the mutual information maximization principle, 
and ADT~\cite{wang2020denoising} focuses on the inaccuracy problem and proposes an adaptive denoising training principle to prune items with large training loss.
Different from these methods, WSLRec tackles the incompleteness and inaccuracy problem jointly: Both extra supervisions and the original ones are regarded as weak supervisions in a unified manner and the proposed training approach does not depend on extra contextual information or specific model architecture. 

To summarize, our main contributions in this paper are:
\begin{itemize}
    \item We propose WSLRec, a novel model-agnostic training approach for resolving the incompleteness and inaccuracy problem in neural sequential recommendation models.
    \item Taking two model-free methods BR and ItemCF as examples, we investigate how to mine extra supervision signals for training sequential recommendation model.
    \item We evaluate WSLRec on both benchmark datasets and online A/B tests. The results show remarkable performance gains and verifies the effectiveness of WLSRec.
\end{itemize}

\section{Related Work}

%In this section, we review the related literature.
In this section, We only review the work mostly related to ours and refer the interested reader to the survey in~\cite{su2009survey,zhang2019deep}.

\subsection{Neural Sequential Recommendation Models}\label{sec:reviewma}

Sequential recommendation is usually formulated as a multiclass classification problem~\cite{hidasi2015session,kang2018self,zhu2018learning} or a multilabel classification problem~\cite{jain2016extreme,zhuo2020learning} to distinguish relevant items (i.e., positive labels) from irrelevant ones (i.e., negative labels).
Choosing the next one or next several items in future behaviors as positive labels directly is almost the standard way in previous work,
while negative labels are sampled randomly to reduce the computational complexity in training.
Difference choices of the distribution for sampling and the loss function for optimizing leads to various methods including sampled softmax~\cite{jean2015using}, noise contrastive estimation~\cite{gutmann2010noise,mnih2013learning} and negative sampling~\cite{Mikolov2013distributed}.
There exist little work discussing how to mining extra positive labels.

Designing advanced model architecture has received lots of attentions, which can be categorized roughly into two parts: learning better user/item representations (i.e., embedding vectors) and learning better matching function for measuring their relevance.
The former usually adopts the dot product or the cosine similarity of embedding vectors as the matching function to measure the user-item relevance, such that generating the top-$k$ recommendation set corresponds to maximum inner product search (MIPS) and can be implemented efficiently in practice~\cite{JDH17}.
%{\bf Representation Learning}.
%One main focus of previous work is about how to better represent the behavior sequence, and a lot of advanced model architecture is introduced for this purpose.
To name a few, GRU4Rec~\cite{hidasi2015session} proposes to use a gated recursive unit (GRU) based recurrent network,
SASRec~\cite{kang2018self} proposes to use a multi-head attention mechanism and BERT4Rec~\cite{sun2019bert4rec} proposes to use a bi-directional encoder based on Transformer~\cite{vaswani2017attention}.
Another line of work focuses on the diversity of user preferences and proposes to use multiple embedding vectors for representing the multiple interest hidden in the behavior sequence~\cite{li2019multi,cen2020controllable,ma2020disentangled}.
%The commonality of these methods is that they usually adopt the dot product or the cosine similarity of embedding vectors as the matching function to measure the user-item relevance.
%As a result, generating the top-$k$ recommendation set corresponds to maximum inner product search (MIPS) and can be implemented efficiently in practice~\cite{JDH17}.
%{\bf Matching Function Learning}.
The latter focuses on designing complex matching functions other than the dot product to better measure the user-item relevance.
For example, NeuCF~\cite{he2017neural} proposes to use a neural network directly as the matching function, Wide\&Deep~\cite{cheng2016wide} proposes to combine a linear model with a neural network while DeepFM~\cite{guo2017deepfm} proposes to replace the linear model in Wide\&Deep with a factorization machine.
However, generating the top-$k$ recommendation for these models requires computing the matching function for every item, which is tremendous in practice.
To solve this problem, tree models like TDMs~\cite{zhu2018learning,zhu2019joint,zhuo2020learning} propose to learn the neural matching function jointly with a tree structure over the item set, and the (approximate) top-$k$ recommendation set can be generated by performing beam search on tree, which achieves logarithmic time complexity.
WSLRec does not introduce additional constraints on model architecture and thus can be applied to these advanced models as well.
In this paper, we choose several benchmark models in offline experiments to verify this.

%\subsubsection{Training Approaches}

\subsection{Weakly Supervised Learning}

Learning from weakly supervisions~\cite{liao2005logistic,natarajan2013learning,zhou2018brief} has been studied extensively in computer vision~\cite{sun2017revisiting,han2018co,mahajan2018exploring} and natural language processing~\cite{medlock2007weakly,meng2018weakly,wang2019clinical}.
The most relevant paradigm to ours is weakly supervised pre-training, in which a model is pre-trained on a large dataset with weak supervisions while fine-tuned on the target task with a clean supervised dataset~\cite{mahajan2018exploring,girshick2014rich,sun2017revisiting}.
However, this cannot be applied to recommendation problems directly, since it is hard to collect even a small clean dataset that contains explicit feedback for relevance between users and items.
WSLRec also adopts a pre-training/fine-tuning framework, but it gets rid of the limit of clean data by leveraging the top-$k$ mining to filter weak supervisions and build reliable supervised signals for fine-tuning adaptively.

\section{Preliminary}

%In this section, we formalize the definition of sequential recommendations and introduce corresponding model architecture. 

\subsection{Problem Formulation} \label{sec:preliminaryprob}

\iffalse
\begin{itemize}
    \item $\mathcal{B}_{u,t}^{(BR)}$
    \item $\mathcal{B}_{u,t}^{(ItemCF)}$
    \item $\mathcal{Y}_{u,t}$
    \item $\mathcal{B}_{u,t}^{(Pre)} \setminus \left( \mathcal{B}_{u,t}^{(BR)} \bigcup \mathcal{B}_{u,t}^{(ItemCF)} \bigcup \mathcal{Y}_{u,t} \right)$
\end{itemize}
\fi

Notations used in the rest of paper are summarized in Table~\ref{tab:notations}. 
Let $\mathcal{V}$ and $\mathcal{U}$ denote the item set and the user set, respectively.
Following previous literature~\cite{hidasi2015session,zhu2018learning,kang2018self,ma2020disentangled,zhou2020s3,cen2020controllable} which focuses on modeling the sequential nature in recommendations, we only consider users' behaviors while leave aside other contextual or content information in this paper.
The original implicit feedback data is thus formulated sequentially: for each user $u \in \mathcal{U}$, we sort all items the user has interacted with in ascending order of the timestamp.
This produces the user behavior sequence $S_u = (v_{u,1},...,v_{u,n_u})$, in which $v_{u,t} \in \mathcal{V}$ is the $t$-th interacted item, $t$ is the sequence order index and $n_u$ is the length of $S_u$.
Given a certain index $t$, $S_u$ can be divided into two parts: the history behavior sequence $X_{u,t}=(v_{u,1},...,v_{u,t})$ and the future behavior sequence $Y_{u,t}=(v_{u,t+1},...,v_{u,n_u})$.
With a little abuse of notations, we use $\mathcal{Y}_u^t=\{v_{u,t+1},...,v_{u,n_u}\}$ to denote the set of future behaviors.

The task of sequential recommendation is to predict the most relevant items to $u$ given the history behaviors $X_{u,t}$ as the input.
This is usually formulated as a top-$k$ selection problem, i.e.,

\begin{equation}\label{eq:topk}
    \mathcal{B}_{u,t} = \argTopk_{v \in \mathcal{V}} p_{\theta}(v|X_{u,t}),
\end{equation}
where $p_{\theta}(v|X_{u,t})$ is the estimated conditional distribution with trainable parameters $\theta$ for measuring the intrinsic relevance of $u$ to $\mathcal{V}$ after given the history behaviors $X_{u,t}$, and $\mathcal{B}_{u,t}$ is a set of size $k$ that contains top-$k$ items with respect to $p_{\theta}(v|X_{u,t})$.

In neural based recommendation models, $p_{\theta}(v|X_{u,t})$ is modeled by a deep neural network $f_{\theta}(X_{u,t},v)$, i.e.,
\begin{equation}\label{eq:pformulate}
    p_{\theta}(v|X_{u,t}) = \frac{\exp \left(f_{\theta}(X_{u,t},v)\right)}{\sum_{v' \in \mathcal{V}} \exp \left(f_{\theta}(X_{u,t},v')\right)}.
\end{equation}

Learning such a deep model $f_{\theta}(X_{u,t},v)$ is thus regarded as a multiclass classification problem, in which $\theta$ is trained to minimize the negative log likelihood $-\log p_{\theta}(v|X_{u,t})$ over the training data, 
\begin{equation}\label{eq:obj}
    \theta^* = \argmin_{\theta} \sum_{u} \sum_{t} \sum_{v \in \mathcal{P}_{u,t}} - \log p_{\theta}(v|X_{u,t}),
\end{equation}
where $\mathcal{P}_{u,t}$ is the set of relevant items to $u$ after history behaviors $X_{u,t}$.
Most previous literature assumes $\mathcal{P}_{u,t}=\{v_{u,t+1}\}$, i.e., only the next immediate item $v_{u,t+1}$ is relevant to $u$.

In practice, since $\mathcal{V}$ is usually extremely large, sub-sampling is widely adopted to reduce the computational complexity of computing the denominator of Eq. (\ref{eq:pformulate}) .
Following~\cite{covington2016deep,li2019multi,cen2020controllable}, we use the sampled softmax loss~\cite{jean2015using} and replace $\log p_{\theta}(v|X_{u,t})$ in Eq. (\ref{eq:obj}) with 
\begin{equation}\label{eq:sampledsoftmax}
    \tilde{f}_{\theta}(X_{u,t},v) - \log \left( \tilde{f}_{\theta}(X_{u,t},v) + \sum_{v' \in \mathcal{N}_{u,t}} \exp (\tilde{f}_{\theta}(X_{u,t},v')) \right).
\end{equation}
In Eq. (\ref{eq:sampledsoftmax}), $\mathcal{N}_{u,t}$ is the set of irrelevant items, which are sampled randomly from $\mathcal{V}$ according to the proposal distribution $Q: \mathcal{V} \to \mathbb{R}$ such that its size satisfies $|\mathcal{N}_{u,t}| \ll |\mathcal{V}|$ and $\tilde{f}_{\theta}(X_{u,t},v)=f_{\theta}(X_{u,t},v)-\log Q(v)$ is weighted according to $Q$.

\begin{table}[]
    \centering
    \caption{Notations.}
    \scalebox{0.96}{
    \begin{tabular}{c l}
    \toprule
        {\bf Notation} & {\bf Description} \\
    \midrule
        $\mathcal{U}$, $\mathcal{V}$ & the set of users and items \\
        $u$, $v$ & a user and an item \\
        $S_u$ & the behavior sequence of user $u$ \\
        $n_u$ & the length of $S_u$ \\
        $v_{u,t}$ & the $t$-th item in the behavior sequence $S_u$ \\
        $X_{u,t}$ & the history behavior sequence of $u$ till index $t$ \\
        $Y_{u,t}$ & the future behavior sequence of $u$ after index $t$ \\
        $\mathcal{Y}_{u,t}$ & the set of items in $Y_{u,t}$ \\
        $\mathcal{P}_{u,t}$ & the set of relevant items to $u$ given $X_{u,t}$ \\
        $\mathcal{N}_{u,t}$ & the set of irrelevant items to $u$ given $X_{u,t}$ \\
        $\Ev \in \mathbb{R}^{|\mathcal{V}|\times d}$ & the embedding matrix for the item set $\mathcal{V}$ \\
        $\Ev_v \in \mathbb{R}^d$ & the embedding vector for item $v$ \\
        $\mathbb{I}(\cdot)$ & the indicator function \\
    \bottomrule
    \end{tabular}
    }
    \label{tab:notations}
\end{table}

\subsection{Model Architecture} \label{sec:preliminarymodel}

As is discussed in Section~\ref{sec:reviewma}, the model architecture $f_{\theta}(X_{u,t}, v)$ is not our main focus and we just implement the state-of-the-art model architecture~\cite{hidasi2015session,kang2018self,li2019multi,cen2020controllable} for evaluating the effectiveness of our proposed WSLRec.
Let $\Ev \in \mathbb{R}^{|\mathcal{V}|\times d}$ denotes the item embedding matrix, the item embedding for $v \in \mathcal{V}$ is thus denoted as $\Ev_v \in \mathbb{R}^d$.
All of these models measure the relevance between $u$ and $v$ by the inner product between the user and item embedding vector, while their difference lies in the user embedding, as detailed below:

 {\bf GRU4Rec}~\cite{hidasi2015session} utilizes a GRU-based RNN over $X_{u,t}$ and use the hidden vector at index $t$, denoted by $\hv(X_{u,t}) \in \mathbb{R}^d$, as the user embedding.
The relevance is thus measured as the inner product between the user embedding and the item embedding, i.e.,
\begin{equation}\label{eq:singlevector}
    f_{\theta}(X_{u,t}, v) = \Ev_v^\top \hv(X_{u,t}).
\end{equation}

{\bf SASRec}~\cite{hidasi2015session} leverages the multi-head self-attention mechanism over $X_{u,t}$ to extract $\hv(X_{u,t})$. Similar to GRU4Rec, it measures the relevance according to Eq. (\ref{eq:singlevector}) as well. 

{\bf MIND}~\cite{li2019multi} generates multiple user embedding vectors instead of a single one to capture different aspects of the user's interest on items, which is denoted as $\hv_1(X_{u,t}),...,\hv_m(X_{u,t}) \in \mathbb{R}^d$.
It utilizes the capsule routing mechanism~\cite{sabour2017dynamic} to cluster history behaviors and generating $m$ user embedding vectors. Unlike GRU4Rec and SASRec, the relevance is measured as
\begin{equation}\label{eq:multivector}
    f_{\theta}(X_{u,t}, v) = \max_{j \in \{1,...,m\}} \Ev_v^\top \hv_j(X_{u,t}).
\end{equation}

{\bf ComiRec-DR}~\cite{cen2020controllable} also leverages the capsule routing mechanism to generate multiple user vectors and measures the relevance as Eq. (\ref{eq:multivector}).
However, it uses a separate bilinear mapping matrix for each pair of low-level capsules and high-level capsules while MIND uses one matrix shared by all the pairs.

For these models, finding top-$K$ set $\mathcal{B}_{u,t}$ defined in Eq. (\ref{eq:topk}) corresponds to maximum inner product search (MIPS), which has been studied extensively~\cite{indyk1998approximate,jegou2010product,ram2012maximum,shrivastava2014asymmetric} and there exists open source libraries implemented efficiently on GPUs, e.g., FAISS~\cite{JDH17}. 

\begin{figure*}[]
    \centering
    \includegraphics[width=0.9\textwidth]{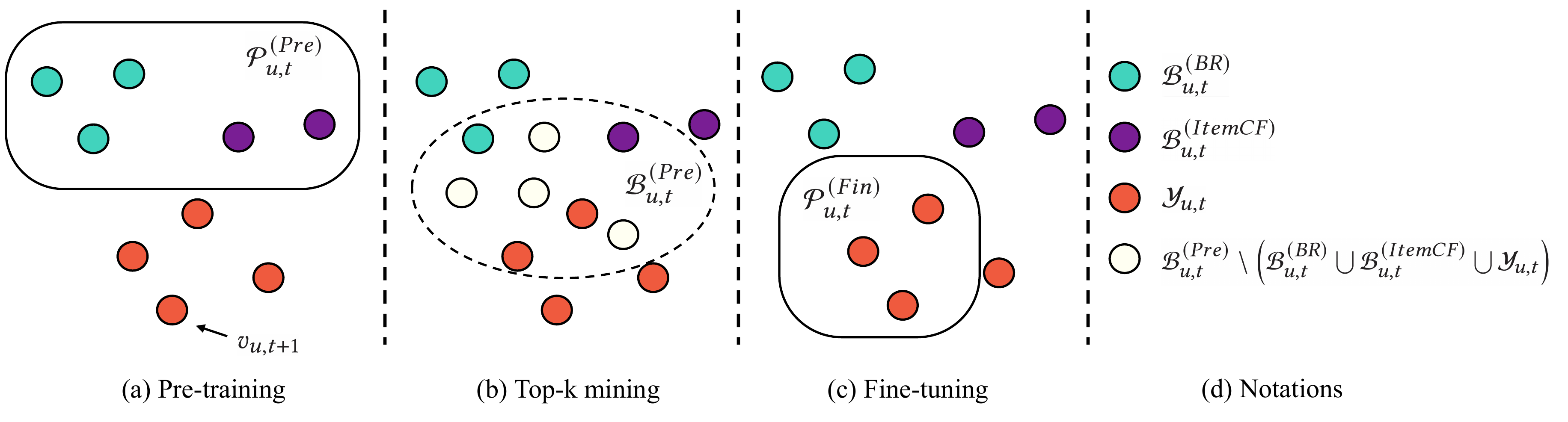}
    \caption{An overview of WSLRec for user $u$ with historical behaviors $X_{u,t}$. (a) The pre-training stage, where the positive label set $\mathcal{P}_{u,t}^{(Pre)}$ is chosen as the union of items generated by BR and ItemCF. (b) The top-$k$ mining stage, where the pre-trained model generates top-$k$ set $\mathcal{B}_{u,t}^{(Pre)}$ (nodes in the dashed circle). (c) The fine-tuning stage, where the positive label set $\mathcal{P}_{u,t}^{(Fin)}$ is chosen as the next immediate item $v_{u,t+1}$ plus the intersection between the top-$k$ set and future behaviors (i.e., $\mathcal{B}_{u,t}^{(Pre)}\bigcap \mathcal{Y}_{u,t}$). (d) The notations, where blue nodes denote items generated by BR, purple nodes denote item generated by ItemCF, red nodes denote items in future behaviors while white nodes denote items in the top-$k$ set but not in BR, ItemCF and future behaviors.}
    \label{fig:framework}
\end{figure*}

\section{Motivation}\label{sec:motivation}

In this section, we provide empirical evidence for supporting the claim that it is {\bf incomplete} and {\bf inaccurate} to regard future behaviors as relevant items (i.e., positive labels in a classification view) for training recommendation models directly.

\subsection{Incompleteness}

In real-world scenarios, only a small subset of items are exposed to a user and the user's behaviors only reflect partial relevance, which results in models trained according to this implicit feedback data may get stuck in local optima.
This is a ubiquitous problem in recommender systems and has been studied extensively as missing-not-at-random (MNAR) in matrix factorization models~\cite{schnabel2016recommendations}.

We investigate its influence on neural sequential recommendation models by comparing the top-$k$ set recommended by models trained according to future behaviors to that recommended by model-free methods which recommend items in a different way from future behavior prediction.
We consider behavioral retargeting (BR)~\cite{yan2009much} and item-based collaborative filtering (ItemCF)~\cite{sarwar2001item} since both of them prevail in practice, though our method proposed in Section~\ref{sec:method} does not limit to BR and ItemCF.

\begin{table}[]
    \centering
     \caption{A comparison of HDR@50 ($\%$) on Taobao User Behavior dataset.}
    \label{tab:demo2}
    \scalebox{0.96}{
    \begin{tabular}{c | c c c c}
    \toprule
        \multirow{2}{*}{Methods} & \multicolumn{4}{|c}{HDR@50} \\
         & BR & ItemCF & GRU4Rec & MIND \\
    \hline
        BR & 0.00 & 91.22 & 83.29 & 76.23 \\
        ItemCF & 91.22 & 0.00 & 91.26 & 88.82 \\
        GRU4Rec & 83.29 & 91.26 & 0.00 & 75.60\\
        MIND & 76.23 & 88.82 & 75.60 & 0.00 \\
    \bottomrule
    \end{tabular}
    }
\end{table}

\subsubsection{Behavioral Retargeting} 
BR recommends an item to users who have already interacted with it in the past, based on the assumption that users tend to interact with items repeatedly.
Given the historical behaviors of user $u$ as $X_{u,t}=\{v_{u,1},...,v_{u,t}\}$, the top-$k$ set generated by BR can be denoted as 
\begin{equation}
    \mathcal{B}_{u,t}^{(BR)} = \{v_{u,i}: t - k < i \leq t\},
\end{equation}
which contains items being the $k$-nearest to $t$.

\subsubsection{Item-based Collaborative Filtering}
ItemCF recommends users with items that are similar to the users' historical behaviors, which relies on the measure of similarity between items.
We adopt the weighted cosine similarity.
Let $\mathcal{U}_{tr}$ denote the set of training users, for any $v, v' \in \mathcal{V}$, the weighted cosine similarity is defined as
\begin{equation}\label{eq:wbcosine}
    \mathrm{sim}(v, v') = \frac{\sum_{u \in \mathcal{U}_{tr}} W_{u,v}W_{u,v'}}{\sqrt{\sum_{u \in \mathcal{U}_{tr}} W_{u,v}^2} \sqrt{\sum_{u \in \mathcal{U}_{tr}} W_{u,v'}^2}},
\end{equation}
where $W_{u,v}$ is defined as
\begin{equation}\label{eq:wuv}
    W_{u,v}=\frac{\sum_{t=1}^{n_u} \mathbb{I}(v_{u,t}=v)}{\log_2 (1+n_u)}.
\end{equation}  
The role of $\log_2 (1+n_u)$ in Eq. (\ref{eq:wuv}) is to prevent $W_{u,v}$ from being too large for users with much more behaviors than others.

Given the historical behaviors $X_{u,t}$, the top-$k$ set generated by ItemCF can be denoted as
\begin{equation}
    \mathcal{B}_{u,t}^{(ItemCF)} = \argTopk_{v \in \mathcal{V}} \max_{v' \in X_{u,t}} \mathrm{sim}(v,v'),
\end{equation}
which selects the $k$ most similar items to any item in historical behaviors according to $\mathrm{sim}(v,v')$.

\subsubsection{Analysis}

For numerical comparison, we adopt the {\it hits difference rate} (HDR)~\cite{zhu2019improving} as the evaluation metric.
For user $u$, let $X_u$ denote the historical behavior sequence and $\mathcal{Y}_u$ denote the ground truth set of future behaviors\footnote{We omit the index $t$ for simplicity in notations.}, HDR is defined as  
\begin{equation}\label{eq:hdr}
    \mathrm{HDR}@k(u;\mathcal{M}_1,\mathcal{M}_2) = \frac{|\mathcal{H}_u^{(\mathcal{M}_1)} \bigcup \mathcal{H}_u^{(\mathcal{M}_2)}| - |\mathcal{H}_u^{(\mathcal{M}_1)} \bigcap \mathcal{H}_u^{(\mathcal{M}_2)}| }{|\mathcal{H}_u^{(\mathcal{M}_1)} \bigcup \mathcal{H}_u^{(\mathcal{M}_2)}|},
\end{equation}
where $\mathcal{M}_1$ and $\mathcal{M}_2$ denotes two recommendation methods, $\mathcal{H}_u^{(\mathcal{M}_1)}=\mathcal{B}_u^{(\mathcal{M}_1)} \bigcap \mathcal{Y}_u$ is the set of items that are in the top-$k$ set generated by $\mathcal{M}_1$ and also the ground truth.
Notice that HDR is set to zero when both $\mathcal{H}_u^{(\mathcal{M}_1)}$ and $\mathcal{H}_u^{(\mathcal{M}_2)}$ are empty.

Table~\ref{tab:demo2} shows the per-user averaged HDR over testing set of Taobao User Behavior dataset, i.e.,
\begin{equation}
    \mathrm{HDR}@k(\mathcal{M}_1,\mathcal{M}_2) = \frac{\sum_{u} \mathrm{HDR}@k(u;\mathcal{M}_1,\mathcal{M}_2)}{\sum_u \mathbb{I}\left(\mathrm{HDR}@k(u;\mathcal{M}_1,\mathcal{M}_2) > 0\right)}.
\end{equation}

We can find out that the HDR between model-free (BR, ItemCF) and model-based (GRU4Rec, MIND) methods is higher than that among model-based methods.  
This implies 
%that even the performance is evaluated on future behaviors, 
the top-$k$ recommendations of BR and ItemCF cover different parts of the test set from models trained according to future behaviors, no matter which model architecture is used.
Inspired by this observation, WSLRec proposed in Section~\ref{sec:method} leverages pre-training on the recommended items provided by these model-free methods to incorporate useful knowledge other than future behavior prediction.

\subsection{Inaccuracy}

Apart from BR and ItemCF, a natural idea for resolving the incompleteness problem is to introduce longer-term future behaviors for training.
However, given the historical behavior sequence, not all future behaviors can be regarded as accurate positive labels, since users' behaviors in longer-term futures may depend on the behaviors in shorter-term futures.

%we can verify that this makes the inaccuracy problem even worse, even the performance is evaluated on future behaviors in the test set.
Table~\ref{tab:demo1} verifies this claim:
The performance improves when a small part (e.g., Next 3) of future behaviors is used for model training, while it deteriorates when longer-term future behaviors are introduced (e.g., Next 10/All).
Inspired by this observation, WSLRec proposed in Section~\ref{sec:method} leverages the top-$k$ mining for finding predictable items from future behaviors according to the pre-trained model, which serve as reliable positive labels for fine-tuning.

\begin{table}[]
    \centering
     \caption{A comparison of recall@20 ($\%$) on Taobao User Behavior dataset for models trained with various parts of future behaviors. Next $c$ denotes $\mathcal{P}_{u,t}=\{v_{u,t+1},...,v_{u,t+c}\}$.}
    \label{tab:demo1}
    \scalebox{0.96}{
    \begin{tabular}{c c c c c}
    \toprule
         & Next 1 & Next 3 & Next 10 & Next All \\
    \midrule
        %GRU4Rec & 8.49 & 8.40 & 6.56 & 4.13 \\
        %MIND & 8.58 & 9.27 & 8.02 & 5.44 \\ 
        GRU4Rec & 5.86 & 5.90 & 4.24 & 2.41 \\
        MIND & 5.84 & 6.03 & 5.48 & 4.19 \\ 
    \bottomrule
    \end{tabular}
    }
\end{table}

%Most neural based recommendation models consider the ``next-one'' strategy. We slightly abuse notations by using $\mathcal{H}_u$ to denote either a set or a sequence sorted in an ascend order of timestamp. 
%These methods usually use user history behaviors $\mathcal{H}_u$ as inputs and use the next behavior as the supervision signals.
%However, as discussed in~\cite{ma2020disentangled}, the next immediate behavior may be irrelevant to the history behaviors.
%On the other hands, using all the future behaviors as supervised signals is questionable.
%Let $t_u^*=\max_{v\in \mathcal{H}_u} t_{u,s}$ denote the 
%\begin{equation}
%    \mathcal{G}_u^{(Future)} = \mathcal{L}_u \setminus \mathcal{H}_u.
%\end{equation}

\section{Method} \label{sec:method}

%In this section, we analyze the importance of $\mathcal{P}_{u,t}$ and $\mathcal{N}_{u,t}$ in learning neural based recommendation models and the drawbacks of the current ``standard'' approach as discussed in Section~\ref{sec:preliminaryprob}.

In this section, we introduce WSLRec for resolving the incompleteness and inaccuracy problem discussed in Section~\ref{sec:motivation}, which can be roughly divided into three stages as is shown in Figure~\ref{fig:framework}.

%This framework does not make any assumptions on the model architecture and thus can be applied to any models.
%At the pretraining stage, the DNNs are trained to optimize the original objective function.
%After the DNNs are pretrained, a query is conducted to get the beam set $\mathcal{B}_u$ for each user $u$.
%To incorporate information from BR and ItemCF, we pre-train the model on top-$k$ set generated by these model-free methods.
%The model is fine-tuned only on future behaviors to make sure that training and evaluation is consistent.
%To resolve the inaccuracy of future behaviors, we utilize the top-$k$ set generated by the pre-trained model as a filter for mining future behaviors and selecting those with high confidence for traing while ignore the others.

\begin{algorithm}
\caption{WSLRec}
\label{alg:WSLRec}
\KwIn{Training set of users $\mathcal{U}_{tr}$, corresponding behavior set $\{S_u:u\in \mathcal{U}_{tr}\}$, model $f_{\theta}(X_{u,t},v)$ with trainable parameter $\theta$.}
\KwOut{The trained parameter $\theta^*$.}
Initializing $\theta$ randomly\\
Computing $\mathrm{v,v'}$ for any $v,v' \in \mathcal{V}$\\
({\bf Pre-training}) Optimizing $\theta$ through minimizing Eq. (\ref{eq:obj}) with $\mathcal{P}_{u,t}^{(Pre)}$ defined in Eq. (\ref{eq:pospre}), and getting the pre-trained parameter $\theta^{(Pre)}$\\
({\bf Top-$k$ Mining}) For each data $X_{u,t}$ with $u \in \mathcal{U}_{tr}$, generating the top-$k$ set 
$$
\mathcal{B}_{u,t}^{(Pre)}=\argTopk_{v \in \mathcal{V}} f_{\theta^{(Pre)}}(X_{u,t}, v).
$$
\\
({\bf Fine-tuning}) Optimizing $\theta$ initialized by $\theta^{(Pre)}$ through minimizing Eq. (\ref{eq:obj}) with $\mathcal{P}_{u,t}^{(Fin)}$ defined in Eq. (\ref{eq:posfin}), and getting the fine-tuned parameter $\theta^*$\\
\end{algorithm}

\subsection{Pre-training}

As is discussed in Section~\ref{sec:motivation}, model-free methods like BR and ItemCF adopt a different way from future behavior prediction to recommend items, and their top-$k$ set hits different parts of the test set compared to neural sequential recommendation models.

Inspired by this phenomenon, we pre-train the model on the top-$k$ set generated by BR and ItemCF.
More specifically, the model is pre-trained by minimizing Eq. (\ref{eq:obj}) with the randomly sampled negative label set $\mathcal{N}_{u,t}$ as shown in Eq. (\ref{eq:sampledsoftmax}) as well.
The only difference to the original training approach is the choice of the positive label set $\mathcal{P}^{(Pre)}_{u,t}$, which is chosen the union of $\mathcal{B}_{u,t}^{(BR)}$ and $\mathcal{B}_{u,t}^{(ItemCF)}$, i.e.,
\begin{equation}\label{eq:pospre}
    \mathcal{P}^{(Pre)}_{u,t} =  \mathcal{B}_{u,t}^{(BR)} \bigcup \mathcal{B}_{u,t}^{(ItemCF)}.
\end{equation}

By doing so, we expect the model can learn useful knowledge for recommending relevant items other than future behavior prediction, and enhance the performance of neural sequential recommendation models through top-$k$ mining and fine-tuning as discussed below.

\subsection{Top-$k$ Mining}

For each $X_{u,t}$ with $u\in \mathcal{U}_{tr}$, the pre-trained model $f_{\theta^{(Pre)}}$ generates the top-$k$ set as
\begin{equation}
    \mathcal{B}_{u,t}^{(Pre)} = \argTopk_{v \in \mathcal{V}} f_{\theta^{(Pre)}}(X_{u,t},v),
\end{equation}
which contains items with the $k$ largest scores with respect to $f_{\theta^{(Pre)}}$ and reflects the generalization property of the model pre-trained on extra supervision signals provided by BR and ItemCF.

\subsection{Fine-tuning}
Similar to other weakly supervised learning approaches in CV~\cite{girshick2014rich,sun2017revisiting}, the fine-tuned model is initialized with the pre-trained parameters to retain the prior knowledge in pre-training.
Since future behaviors are not clean and may deteriorate the performance of neural sequential recommendation models, WSLRec resorts to the top-$k$ set $\mathcal{B}_{u,t}^{(Pre)}$ generated by the pre-trained model as a filter for mining accurate relevant items from future behaviors.
The model is fine-tuned by minimizing Eq. (\ref{eq:obj}) with randomly sampled negative label set in Eq. (\ref{eq:sampledsoftmax}) as well, but the positive label set is chosen as 
\begin{equation}\label{eq:posfin}
    \mathcal{P}_{u,t}^{(Fin)} = \{v_{u,t+1}\} \bigcup \left( \mathcal{B}_{u,t}^{(Pre)} \bigcap \mathcal{Y}_{u,t} \right),
\end{equation}
which degenerates to the set of the next immediate item when the intersection set is empty.
The intuition behind Eq. (\ref{eq:posfin}) is: Apart from the next immediate item, only items that achieve a consensus between weak supervisions used in pre-training and those in the whole future behaviors are regarded as relevant items\footnote{
Since the quality of recommendations is finally measured on future behaviors, we only consider items in future behaviors as potential relevant items in the fine-tuning stage, while investigating other choices is left for future work.}.

\section{Experiments} \label{sec:experiment}

\begin{table}[]
    \centering
    \caption{Statistics of datasets used in experiments.}
    \label{tab:dataset}
    \scalebox{0.96}{
    \begin{tabular}{c c c}
    \toprule
        Dataset  & Amazon Books & Taobao \\
    \midrule
        \# Users & 534,603 & 976,779 \\
        \# Items & 635,522 & 1,708,530 \\
        \# Record & 7,490,071 & 85,384,110 \\
        %Avg. Length & 14.5 & 97.3 \\
    \bottomrule
    \end{tabular}
    }
\end{table}

In this section, we evaluate the performance of WSLRec on both two benchmark datasets and online A/B tests. 
Besides, we also want to answer the following questions:

{\bf RQ1: } Does WSLRec incorporate useful information from extra weak supervisions and thus resolve the incompleteness problem?

{\bf RQ2: } Does WSLRec resolve the inaccuracy problem?

{\bf RQ3: } How do various choices of weak supervision in the pre-training stage affect the final performance of WSLRec?

{\bf RQ4: } Is the top-$k$ mining or fine-tuning necessary for WSLRec?

\subsection{Experimental Setup}

We follow the setup and the code\footnote{\url{https://github.com/THUDM/ComiRec}} of \cite{cen2020controllable}.
This makes the result on Taobao in Table \ref{tab:sota} comparable to Table 3 in \cite{cen2020controllable} except NDCG though the definition in ours and \cite{cen2020controllable} is the same, since their code adopts a different implementation and we fix it.
Our code for reproducing the experiment results will be available upon acceptance.

\subsubsection{Dataset}

We use two large-scale benchmark datasets:
\begin{itemize}
    \item {\bf Amazon Books}~\cite{mcauley2015image,he2016ups} collects users' reviews and ratings on {\it books} of Amazon. We only keep user-book interactions with ratings of four or higher, which differs from \cite{cen2020controllable}.
    \item {\bf Taobao User Behavior}~\cite{zhu2018learning} collects users' behaviors in Taobao with four behavior types: click, purchase, adding item to the shopping cart, and item favoring. We only keep user-item interactions of click.
\end{itemize}

Each record in these datasets is organized in the format of user-item interaction with the timestamp. 
We filter out users that interact with less than 5 items and items that interact with less than 5 users.
For each user $u$, we sort the user's interacted items in an ascend order according to the timestamp and thus produce the user's behavior sequence $S_u$, which contains $n_u \geq 5$ items.
Statistics of these datasets after pre-processing are summarized in Table~\ref{tab:dataset}.

We split users randomly into training ($\mathcal{U}_{tr}$), validation ($\mathcal{U}_{va}$) and test sets ($\mathcal{U}_{te}$) with the proportion of 8:1:1.
For each training user $u$, since $n_u \geq 5$, we traverse the index $t$ from $4$ to $n_u-1$ to guarantee both the historical behavior sequence $X_{u,t}$ and the future behavior sequence $Y_{u,t}$ are not empty.
This generates $n_u-4$ training instances, i.e., $\{(X_{u,t}, Y_{u,t})\}_{t=4}^{n_u-1}$, for user $u$. 
To make sure $X_{u,t}$ in a training batch having equal lengths, we further truncate $X_{u,t}$ in Amazon Books dataset at length 20, while that in Taobao User Behavior dataset at length 50.
While for the user $u$ in the validation or test set, we regard the first 80\% of $S_u$ as the historical behavior sequence $X_u$ ($t$ is omitted for notation simplicity), and regard the set of items in the latter 20\%, denoted by $\mathcal{Y}_u$, as the ground truth for evaluating the retrieval performance.

\begin{table*}[h]
    \centering
    \addtolength{\tabcolsep}{-2pt}
    \caption{An overall performance comparison of metrics@$k$ (\%) with $k=20, 50$ on the benchmark datasets. We repeat the experiment with random initialization for 5 times and report the mean and the standard deviation.}
    \label{tab:sota}
\scalebox{0.96}{
    \begin{tabular}{l | l l | c c c c c | c c c c c }
    \toprule 
        \multirow{2}{3.5em}{Datasets} & \multirow{2}{*}{Methods} & \multirow{2}{*}{Variants} & \multicolumn{5}{c|}{Metrics@20} & \multicolumn{5}{c}{Metrics@50} \\
        & & & Prec & Recall & F1 & NDCG & HitRate & Prec & Recall & F1 & NDCG & HitRate \\
    \hline
\multirow{8}{3.5em}{Amazon Books} 
& \multirow{2}{*}{GRU4Rec} & Original & 0.46 $_{\pm .01}$ & 3.52 $_{\pm .09}$ & 1.99 $_{\pm .05}$ & 3.52 $_{\pm .09}$ & 7.46 $_{\pm .14}$ & 0.30 $_{\pm .01}$ & 5.47 $_{\pm .10}$ & 2.89 $_{\pm .05}$ & 4.26 $_{\pm .09}$ & 11.39 $_{\pm .18}$ \\
& & WSLRec & 1.11 $_{\pm .01}$ & 8.79 $_{\pm .06}$ & 4.95 $_{\pm .03}$ & 7.51 $_{\pm .06}$ & 16.61 $_{\pm .09}$ & 0.69 $_{\pm .01}$ & 12.66 $_{\pm .04}$ & 6.67 $_{\pm .03}$ & 8.79 $_{\pm .05}$ & 23.25 $_{\pm .08}$ \\
\cline{2-13} 
& \multirow{2}{*}{SASRec} & Original & 0.56 $_{\pm .00}$ & 4.62 $_{\pm .06}$ & 2.59 $_{\pm .03}$ & 4.25 $_{\pm .03}$ & 9.29 $_{\pm .11}$ & 0.35 $_{\pm .00}$ & 6.81 $_{\pm .05}$ & 3.58 $_{\pm .03}$ & 5.05 $_{\pm .03}$ & 13.50 $_{\pm .10}$ \\
& & WSLRec & 1.18 $_{\pm .01}$ & 9.34 $_{\pm .05}$ & 5.26 $_{\pm .03}$ & 8.04 $_{\pm .06}$ & 17.45 $_{\pm .14}$ & 0.71 $_{\pm .01}$ & 13.24 $_{\pm .10}$ & 6.98 $_{\pm .06}$ & 9.28 $_{\pm .06}$ & 23.95 $_{\pm .15}$ \\
\cline{2-13} 
& \multirow{2}{*}{MIND} & Original & 0.81 $_{\pm .01}$ & 6.48 $_{\pm .04}$ & 3.65 $_{\pm .02}$ & 6.42 $_{\pm .03}$ & 12.92 $_{\pm .10}$ & 0.49 $_{\pm .00}$ & 9.05 $_{\pm .05}$ & 4.77 $_{\pm .03}$ & 7.33 $_{\pm .04}$ & 17.73 $_{\pm .14}$ \\
& & WSLRec & \textbf{1.29} $_{\pm .01}$ & \textbf{10.22} $_{\pm .04}$ & \textbf{5.75} $_{\pm .02}$ & \textbf{8.94} $_{\pm .04}$ & \textbf{19.14} $_{\pm .06}$ & \textbf{0.77} $_{\pm .01}$ & \textbf{14.30} $_{\pm .08}$ & \textbf{7.53} $_{\pm .04}$ & \textbf{10.23} $_{\pm .05}$ & \textbf{25.94} $_{\pm .11}$ \\
\cline{2-13} 
& \multirow{2}{*}{ComiRec} & Original & 0.62 $_{\pm .01}$ & 4.99 $_{\pm .07}$ & 2.81 $_{\pm .04}$ & 5.35 $_{\pm .10}$ & 10.03 $_{\pm .12}$ & 0.38 $_{\pm .00}$ & 7.07 $_{\pm .12}$ & 3.72 $_{\pm .06}$ & 6.10 $_{\pm .10}$ & 14.07 $_{\pm .17}$ \\
& & WSLRec & 0.83 $_{\pm .01}$ & 6.77 $_{\pm .11}$ & 3.80 $_{\pm .06}$ & 6.42 $_{\pm .13}$ & 12.99 $_{\pm .21}$ & 0.49 $_{\pm .01}$ & 9.60 $_{\pm .11}$ & 5.05 $_{\pm .06}$ & 7.39 $_{\pm .13}$ & 18.13 $_{\pm .22}$ \\
\hline \hline
\multirow{8}{3.5em}{Taobao User Behavior} 
& \multirow{2}{*}{GRU4Rec} & Original & 3.95 $_{\pm .02}$ & 5.85 $_{\pm .02}$ & 4.90 $_{\pm .02}$ & 18.48 $_{\pm .04}$ & 35.55 $_{\pm .14}$ & 2.37 $_{\pm .02}$ & 8.49 $_{\pm .06}$ & 5.43 $_{\pm .04}$ & 19.99 $_{\pm .05}$ & 46.00 $_{\pm .19}$ \\
& & WSLRec & 4.96 $_{\pm .08}$ & 6.73 $_{\pm .18}$ & 5.84 $_{\pm .13}$ & 21.47 $_{\pm .39}$ & 40.04 $_{\pm .72}$ & 2.96 $_{\pm .06}$ & 9.72 $_{\pm .29}$ & 6.34 $_{\pm .17}$ & 22.88 $_{\pm .41}$ & 51.04 $_{\pm .92}$ \\
\cline{2-13}
& \multirow{2}{*}{SASRec} & Original & 4.22 $_{\pm .03}$ & 6.11 $_{\pm .03}$ & 5.16 $_{\pm .03}$ & 20.25 $_{\pm .11}$ & 37.31 $_{\pm .19}$ & 2.49 $_{\pm .02}$ & 8.65 $_{\pm .03}$ & 5.57 $_{\pm .03}$ & 21.65 $_{\pm .11}$ & 47.65 $_{\pm .29}$ \\
& & WSLRec & 4.77 $_{\pm .01}$ & 6.67 $_{\pm .02}$ & 5.72 $_{\pm .02}$ & 21.14 $_{\pm .09}$ & 39.83 $_{\pm .12}$ & 2.86 $_{\pm .01}$ & 9.63 $_{\pm .03}$ & 6.24 $_{\pm .02}$ & 22.64 $_{\pm .08}$ & 51.10 $_{\pm .09}$ \\
\cline{2-13}
& \multirow{2}{*}{MIND} & Original & 4.23 $_{\pm .00}$ & 5.82 $_{\pm .02}$ & 5.03 $_{\pm .01}$ & 18.76 $_{\pm .06}$ & 36.20 $_{\pm .07}$ & 2.58 $_{\pm .01}$ & 8.58 $_{\pm .02}$ & 5.58 $_{\pm .01}$ & 20.37 $_{\pm .06}$ & 47.41 $_{\pm .09}$ \\
& & WSLRec & \textbf{6.05} $_{\pm .01}$ & \textbf{8.34} $_{\pm .03}$ & \textbf{7.19} $_{\pm .02}$ & \textbf{25.28} $_{\pm .04}$ & \textbf{46.38} $_{\pm .06}$ & \textbf{3.47} $_{\pm .01}$ & \textbf{11.57} $_{\pm .04}$ & \textbf{7.52} $_{\pm .02}$ & \textbf{26.38} $_{\pm .08}$ & 56.65 $_{\pm .10}$ \\
\cline{2-13}
& \multirow{2}{*}{ComiRec} & Original & 5.07 $_{\pm .06}$ & 7.28 $_{\pm .08}$ & 6.18 $_{\pm .07}$ & 23.03 $_{\pm .23}$ & 42.76 $_{\pm .38}$ & 3.00 $_{\pm .04}$ & 10.38 $_{\pm .11}$ & 6.69 $_{\pm .08}$ & 24.40 $_{\pm .24}$ & 53.76 $_{\pm .45}$ \\
& & WSLRec & 5.55 $_{\pm .30}$ & 7.69 $_{\pm .39}$ & 6.62 $_{\pm .35}$ & 24.77 $_{\pm .86}$ & 45.40 $_{\pm .47}$ & 3.28 $_{\pm .19}$ & 11.05 $_{\pm .58}$ & 7.16 $_{\pm .38}$ & 26.10 $_{\pm .85}$ & \textbf{56.69} $_{\pm .62}$ \\
    \bottomrule
    \end{tabular}
}
\end{table*}

\subsubsection{Base Models and Implementation Details} 
In experiments we consider four base models: GRU4Rec~\cite{hidasi2015session}, SASRec~\cite{kang2018self}, MIND~\cite{li2019multi} and ComiRec-DR~\cite{cen2020controllable}.
For each historical behavior sequence, the former two models produce one embedding vector, while the latter twos produce multiple embedding vectors.

The embedding dimension $d$ of the above models is set to 64.
For MIND and ComiRec-DR, the number of multiple embedding vectors is set to 4 and the number of dynamic routing iterations is set to 3.
The batch size is set to 256.
The number of negative labels for sampled softmax loss is set to 10 per instance, and training instances in the same batch shares these negative labels.
Adam optimizer~\cite{kingma2014adam} with a learning rate 0.001 is adopted to train the model for up to 1 million iterations and early stopping on recall@50 on the validate dataset is adopted to prevent overfitting. 

If without further statement, WSLRec uses the top-20 set of BR for pre-training on Taobao User Behavior dataset while that of both BR and ItemCF for pre-training on Amazon Books dataset.
The reason for this setting is analyzed in Section~\ref{sec:rq4}.
WSLRec follows Eq. (\ref{eq:posfin}) for fine-tuning on both datasets.

\subsubsection{Evaluation Metrics}
We evaluate the performance of top-$k$ recommendations on the test set $\mathcal{U}_{te}$ by the following metrics\footnote{We follow the per-user average definition in~\cite{zhu2018learning,cen2020controllable} , which is different from the macro/micro average definition. We refer interested readers to~\cite{wu2017unified} for more details.}:
\begin{itemize}
    \item Precision, recall and F-measure, i.e.,
    \begin{equation}\label{eq:prec}
        \mathrm{Precision}@k = \frac{1}{|\mathcal{U}_{te}|} \sum_{u \in \mathcal{U}_{te}} \frac{|\mathcal{B}_u \bigcap \mathcal{Y}_u|}{k},
    \end{equation}
    \begin{equation}
        \mathrm{Recall}@k = \frac{1}{|\mathcal{U}_{te}|} \sum_{u \in \mathcal{U}_{te}} \frac{|\mathcal{B}_u \bigcap \mathcal{Y}_u|}{|\mathcal{Y}_u|},
    \end{equation}
    \begin{equation}\label{eq:fmeasure}
        \mathrm{F-Measure}@k =  \frac{1}{|\mathcal{U}_{te}|} \sum_{u \in \mathcal{U}_{te}} \frac{2 \cdot |\mathcal{B}_u \bigcap \mathcal{Y}_u| }{k + |\mathcal{Y}_u|}.
    \end{equation}
    \item Hit rate, i.e.,
    \begin{equation}\label{eq:hit}
        \mathrm{Hit}@k = \frac{1}{|\mathcal{U}_{te}|} \sum_{u \in \mathcal{U}_{te}} \mathbb{I}(|\mathcal{B}_u \bigcap \mathcal{Y}_u| > 0).
    \end{equation}
    \item Normalized discounted cumulative gain (NDCG), i.e.,
    \begin{equation}\label{eq:hit}
        \mathrm{NDCG}@k = \frac{1}{|\mathcal{U}_{te}|} \sum_{u \in \mathcal{U}_{te}} \sum_{i=1}^{k} \frac{I(v_i \in \mathcal{Y}_u)}{\log_2 (i+1)},
    \end{equation}
    where $v_i$ denotes the $i$-th largest items in $\mathcal{B}_u$.
\end{itemize} 

\iffalse
Besides, to give a thorough comparison, we also consider evaluation metrics for measuring the diversity and novelty of the candidate set generated by different methods.
Following~\cite{bradley2001improving,ziegler2005improving,zhang2008avoiding,cen2020controllable}, the top-$k$ diversity (Diversity$@k$) is defined as the intra-set dissimilarity of $\mathcal{B}_u$, where the dissimilarity is measured by difference in categories between items, i.e., 
\begin{equation}
    \mathrm{Diversity}@k = \frac{1}{|\mathcal{U}|}\sum_{u \in \mathcal{U}} \sum_{v,v' \in \mathcal{B}_u, v \neq v'} \frac{\mathbb{I}(\mathrm{Cate}(v)\neq \mathrm{Cate}(v'))}{k(k-1)/2}.
\end{equation}

Following the idea in~\cite{castells2011novelty,vargas2011rank,zhu2018learning}, the top-$k$ novelty (Novelty$@k$) as the inter-set dissimilarity between $\mathcal{B}_u$ and $\mathcal{H}_u$, where $\mathcal{H}_u$ denotes the set of items that have interactions with user $u$ before recommending, i.e.,
\begin{equation}
    \mathrm{Novelty}@k = \frac{1}{|\mathcal{U}|}\sum_{u \in \mathcal{U}} \frac{|\mathrm{Cate}(\mathcal{B}_u) \setminus \mathrm{Cate}(\mathcal{H}_u)|}{|\mathrm{Cate}(\mathcal{B}_u)|},
\end{equation}
.where $\mathrm{Cate}(\mathcal{B}_u)=\{\mathrm{Cate}(v):v \in \mathcal{B}_u\}$ denotes the set of categories for the candidate set $\mathcal{B}_u$.
\fi

\subsection{Overall Performance Comparison}

Table~\ref{tab:sota} compares the performance between our proposed training framework (i.e., {\bf WSLRec}) and the standard training approach discussed in Section~\ref{sec:preliminaryprob} (i.e., {\bf Original}) which regards the next immediate items as the only positive label.
As is shown, with few bells and whistles, all the evaluation metrics on all four models are improved by a significantly large margin on both benchmarks.
More specifically, without modifying model architecture or introducing additional training data outside the behavior sequence, WSLRec achieves 58.0\% and 11.5\% relative recall@50 lift on Amazon Books and Taobao User Behavior dataset, respectively.

\subsection{Further Analysis}

\subsubsection{Analysis on Hits Difference Rate (RQ1)}

\begin{table}[]
    \centering
     \caption{A comparison of HDR@50 ($\%$) to BR and ItemCF on Taobao User Behavior dataset.}
    \label{tab:overlap}
    \scalebox{0.96}{
    \begin{tabular}{l c c c}
    \toprule
        \multirow{2}{*}{Methods} & \multirow{2}{*}{Variants} & \multicolumn{2}{c}{HDR@50} \\
        \cmidrule{3-4}
         & & BR & ItemCF \\
    \midrule
        \multirow{2}{*}{GRU4Rec} & Original & 83.29 & 91.36 \\
        & WSLRec & 80.60 & 86.48 \\
        \multirow{2}{*}{MIND} & Original & 76.23 & 88.82 \\
        & WSLRec & 68.43  & 86.29 \\
    \bottomrule
    \end{tabular}
    }
\end{table}

To answer RQ1, we analyze the hits difference rate defined in Eq. (\ref{eq:hdr}) on both the standard-trained models and WSLRec-trained models.
Table~\ref{tab:overlap} shows the experimental results on the Taobao User Behavior dataset. 
We can find out that the WSLRec-trained models (no matter GRU4Rec or MIND) have lower hits difference rate to BR and ItemCF than the standard-trained models, which implies the top-$k$ recommendations of WSLRec-trained models are more similar to BR and ItemCF than the standard-trained ones.
This verifies our claims that WSLRec-trained models maintain useful knowledge provided by BR and ItemCF, which is different from future behavior prediction.
Besides, the HDR to BR is lower than that to ItemCF, which coincides with the results in Table~\ref{tab:pre-training} that weak supervisions from BR contribute more than those from ItemCF on the Taobao User Behavior dataset.

\begin{table}[]
    \centering
    \caption{An ablation study of the top-$k$ mining and fine-tuning with metrics@20 (\%) on Amazon Books dataset.}
    \label{tab:mining}
    \scalebox{0.96}{
    \begin{tabular}{l c c c c c c}
    \toprule
        \multirow{2}{*}{Methods} & \multirow{2}{*}{Variants} & \multicolumn{5}{c}{Metrics@20} \\
        \cmidrule{3-7}
        & &  Prec & Recall & F1 & NDCG & HitRate \\
    \midrule
        \multirow{4}{*}{GRU4Rec} 
        & Original & 0.45 & 3.45 & 1.95 & 3.42 & 7.32 \\
        %& Original-3 & 0.56 & 4.28 & 2.42 & 4.09 & 9.00 \\
        & Ensemble & 0.91 & 7.57 & 4.24 & 6.46 & 13.88 \\
        & Fine-tune & 1.04 & 8.40 & 4.72 & 6.89 & 15.90 \\
        & WSLRec  & \bf{1.10} & \bf{8.73} & \bf{4.91} & \bf{7.46} & \bf{16.50} \\
        \midrule
        \multirow{4}{*}{MIND} 
        & Original & 0.81 & 6.47 & 3.64 & 6.45 & 12.92 \\
        %& Original-3 & 0.84 & 6.58 & 3.71 & 6.53 & 13.02 \\
        & Ensemble & 0.97 & 8.03 & 4.50 & 7.15 & 15.22 \\
        & Fine-tune & 1.18 & 9.69 & 5.43 & 7.94 & 18.13 \\
        & WSLRec & \bf{1.29} & \bf{10.24} & \bf{5.76} & \bf{8.97} & \bf{19.19} \\
    \bottomrule
    \end{tabular}
    }
\end{table}

\begin{table*}[]
    \centering
    \caption{A comparison of pre-training and fine-tuning metrics@20 (\%) of various pre-training tasks on benchmark datasets.}
    \label{tab:pre-training}
    \scalebox{0.96}{
    \begin{tabular}{l | l l | c c c | c c c}
    \toprule
        \multirow{2}{4em}{Dataset} & \multirow{2}{4em}{Methods} & \multirow{2}{4em}{Tasks} & \multicolumn{3}{c|}{Pre-training Metrics@20} & \multicolumn{3}{c}{Fine-tuning Metrics@20} \\
        \cline{4-9}
        & & & Recall & NDCG & HitRate & Recall & NDCG & HitRate \\
        \hline
        \multirow{10}{5em}{Amazon Books}
        & \multirow{5}{*}{GRU4Rec} 
          & ItemCF & 4.23 & 3.16 & 7.37 & 7.14 & 6.22 & 13.32 \\
        & & BR & 2.01 & 1.91 & 4.70 & 4.46 & 4.16 & 9.20 \\
        & & Original & 3.45 & 3.42 & 7.32 & 4.29 & 4.28 & 9.05 \\
        & & ItemCF $\bigcup$ BR & \bf{6.04} & \bf{4.79} & \bf{11.42} & \bf{8.73} & \bf{7.46} & \bf{16.50} \\
        & & ItemCF $\bigcup$ BR $\bigcup$ Original & 5.76 & 4.52 & 10.75 & 7.15 & 6.35 & 13.80 \\
        \cline{2-9}
        & \multirow{5}{*}{MIND} 
          & ItemCF & 7.53 & 5.97 & 13.14 & 9.99 & 8.89 & 18.49 \\
        & & BR & 4.00 & 2.85 & 8.34 & 7.23 & 5.37 & 14.18 \\
        & & Original & 6.47 & 6.45 & 12.92 & 7.32 & 7.38 & 14.52 \\
        & & ItemCF $\bigcup$ BR & 8.23 & 5.93 & 14.74 & \bf{10.24} & \bf{8.97} & \bf{19.19} \\
        & & ItemCF $\bigcup$ BR $\bigcup$ Original & \bf{8.95} & \bf{7.04} & \bf{16.56} & 9.68 & 8.89 & 18.34 \\
        \hline \hline
        \multirow{10}{5em}{Taobao User Behavior}
        & \multirow{5}{*}{GRU4Rec} 
          & ItemCF & 1.54 & 4.59 & 9.10 & 4.44 & 13.19 & 27.02 \\
        & & BR & 5.12 & 18.12 & 34.35 & \bf{6.60} & \bf{21.36} & \bf{39.72} \\
        & & Original & \bf{5.86} & \bf{18.43} & \bf{35.47} & 6.30 & 19.45 & 37.94 \\
        & & ItemCF $\bigcup$ BR & 3.22 & 8.99 & 19.76 & 5.59 & 16.92 & 34.14 \\
        & & ItemCF $\bigcup$ BR $\bigcup$ Original & 3.17 & 8.48 & 19.30 & 5.83 & 17.66 & 35.41 \\
        \cline{2-9}
        & \multirow{5}{5em}{MIND} 
          & ItemCF & 3.21 & 9.32 & 19.29 & 5.96 & 18.06 & 35.76 \\
        & & BR & \bf{7.89} & \bf{24.97} & \bf{45.91} & \bf{8.32} & \bf{25.19} & \bf{46.37} \\
        & & Original & 5.84 & 18.71 & 36.16 & 6.49 & 20.76 & 39.50 \\
        & & ItemCF $\bigcup$ BR & 6.89 & 20.05 & 39.12 & 7.42 & 22.43 & 42.54 \\
        & & ItemCF $\bigcup$ BR $\bigcup$ Original & 5.77 & 16.66 & 33.26 & 7.13 & 21.57 & 41.22 \\
    \bottomrule
    \end{tabular}
    }
\end{table*}

\subsubsection{Analysis on Pre-training Tasks (RQ2, RQ3)}\label{sec:rq4}
 
We ablate the choices of different pre-training tasks in detail to investigate their contributions to the final performance of WSLRec.
The results is listed in Table~\ref{tab:pre-training}, where {\bf ItemCF}, {\bf BR} and {\bf Original} denote the positive label set for pre-training is generated from ItemCF, BR and the standard-trained model, and $\bigcup$ denotes the union of set.

RQ2 can be answered by comparing the results in Table~\ref{tab:pre-training} to those in Table~\ref{tab:demo1}.
Since WSLRec fine-tunes the models on the whole future behaviors, we choose the fine-tuning recall@20 in Table~\ref{tab:pre-training} where the pre-training is conducted only on items from future behaviors (denoted as {\bf Original} in {\it Tasks}) to make sure no extra knowledge from BR and ItemCF is introduced and the comparison is fair.
The results are 6.30\% for GRU4Rec and 6.49\% for MIND, which are higher than even the best results in Table~\ref{tab:demo1} (5.71\% for GRU4Rec and 6.03\% for MIND), let alone the results of standard-trained models according to the whole future behaviors(2.41\% for GRU4Rec and 4.19\% for MIND).
This is a direct answer to RQ2.
%We can find that when only leveraging future behaviors for training, WSLRec (i.e., {\bf Original} in {\it Tasks}) achieve much higher recall (i.e., 6.30\% versus 2.41\% for GRU4Rec and 6.49\% versus 4.19\% for MIND) than the standard training approach (i.e., {\bf Next All} in Table~\ref{tab:demo1}).

Table~\ref{tab:pre-training} also provides answers for RQ3:
(1) Fine-tuning metrics are higher than pre-training metrics, which verifies the necessity of fine-tuning again. More importantly, fine-tuning on the standard-trained models also improves the performance (e.g., $5.86\% \to 6.30\%$ w.r.t. recall for GRU4Rec on Taobao User Behavior dataset), which verifies the effectiveness of WSLRec in mining useful information even without BR and ItemCF;
(2) BR contributes less than ItemCF on both pre-training and fine-tuning metrics on Amazon Books dataset, while it performs better on Taobao User Behavior dataset. A possible explanation is that a user may view or click an item repeatedly (i.e., user behaviors on Taobao), but it is less likely for the user to give a rating or write a review to an item multiple times (i.e., book reviews on Amazon);
(3) The best fine-tuning metric is achieved when the pre-trained model does not incorporates supervisions from future behaviors (i.e., no {\bf Original} appears in {\it Tasks}), which implies that keeping different training targets between the pre-training and fine-tuning stage is critical for WSLRec;
(4) Training the model on BR, ItemCF and future behaviors jointly (i.e., the pre-training metrics for ItemCF $\bigcup$ BR $\bigcup$ Original) achieves worse performance than the fine-tuning metrics, which implies that the pre-training and fine-tuning framework is more effective in mining useful information.
\iffalse
\begin{itemize}
    \item Fine-tuning metrics are higher than pre-training metrics, which verifies the necessity of fine-tuning in WSLRec again. More importantly, fine-tuning on the standard-trained models also improves the performance (e.g., $5.86\% \to 6.30\%$ w.r.t. recall for GRU4Rec on Taobao User Behavior dataset), which implies that WSLRec brings improvement even without incorporating extra knowledge from BR and ItemCF. 
    \item BR contributes less than ItemCF on both pre-training and fine-tuning metrics on the Amazon Books dataset, while it performs better on Taobao User Behavior dataset. A possible explanation is that a user may view or click an item repeatedly (i.e., user behaviors on Taobao), but it is less likely for the user to give a rating or write a review to an item multiple times (i.e., book reviews on Amazon). 
    \item The best fine-tuning metric is achieved when the pre-trained model does not incorporates supervisions from future behaviors (i.e., no {\bf Original} appears in {\it Tasks}), which implies that keeping different training targets between the pre-training and fine-tuning stage is critical for WSLRec.
    %\item ItemCF performs much worse on Taobao User Behavior dataset than that on Amazon Books dataset, which implies that the item-to-item similarity may be hard to measure on the Taobao User Behavior dataset. All the items in Amazon Books dataset belong to the category of {\it Books} while items in Taobao User Behavior dataset may belong to various categories like {\it Electronics} and {\it Clothes}.
\end{itemize}
\fi

\subsubsection{Analysis on the Top-$k$ Mining and Fine-tuning (RQ4)}

Apart from {\bf WSLRec} and {\bf Original}, we further consider two strategies for leveraging pre-trained models: 
\begin{itemize}
    %\item {\bf Original-3:} A variant of {\bf Original} which regards the next 3 items in future behaviors as positive labels, which corresponds to {\it Next 3} in Table~\ref{tab:demo2}.
	\item {\bf Ensemble:} The final top-$k$ set merges the top-$a$, top-$b$ and top-$c$ items of BR, ItemCF and the standard-trained model such that $a+b+c=k$. The hyper-parameters $a,b,c$ are tuned on recall@$k$ on the validation set. 
	\item {\bf Fine-tune:} A variant of WSLRec that removes the top-$k$ mining stage and fine-tunes the pre-trained models directly on the next immediate item.
\end{itemize}
 
Table~\ref{tab:mining} shows that {\bf Ensemble} outperforms {\bf Original} and {\bf Fine-tune} outperforms {\bf Ensemble}, which verifies again that BR and ItemCF provide useful knowledge and implies that fine-tuning is more effective in mining useful knowledge from BR and ItemCF than merging the set of recommended items directly.
%This verifies the necessity of the fine-tuning stage. 
Besides, WSLRec further achieves remarkable performance gains compared to Fine-tune, which proves the proposed top-$k$ mining stage is necessary.
Results in Table~\ref{tab:ablation_top_k} shows that choosing $k=50$ makes a good trade-off between performance and fine-tuning cost, which answers why we choose $k=50$ as the default parameter.

\begin{table}[]
    \centering
     \caption{A comparison of recall@20 ($\%$) on Amazon Books with various $k$ in the top-$k$ mining stage.}
    \label{tab:ablation_top_k}
    \scalebox{0.96}{
    \begin{tabular}{c c c c}
    \toprule
        $k$ & 5 & 50 & 500 \\
    \midrule
        GRU4Rec & 8.56 & 8.79 & 8.60 \\
        MIND & 9.85 & 10.22 & 10.24 \\ 
    \bottomrule
    \end{tabular}
    }
\end{table}

\subsection{Online A/B Tests}

We conduct an online A/B test for more than one month in Alibaba display advertising system, which provides billions of advertisement impressions to users every day.
In the testing, the only variable is the training algorithm (i.e., WSLRec or the original one), and WSLRec trained model contributes up to 2.1\% click-through rate (CTR) and 3.2\% revenue per mille (RPM) promotion compared to the original model.
This is a significant improvement and demonstrates the effectiveness of WSLRec.
Now WSLRec has been deployed online and serves the main traffic.

\section{Conclusion and Future Work}

We propose WSLRec for resolving the incompleteness and inaccuracy problems in sequential recommendation models, which relies on pre-training to incorporate extra weak supervisions, the top-$k$ mining for mining reliable positive labels, and fine-tuning for improving the model based on these labels. 
For future work, we'd like to investigate more auxiliary weak supervisions by leveraging contextual information like item attributes.
%Besides, investigating WSLRec based on the formulation of learning to rank (e.g.,~\cite{burges2005learning}) is also an interesting direction.

\bibliographystyle{ACM-Reference-Format}
\bibliography{wslrec}

\appendix

\end{document}